\date{April 27, 1998}
\documentstyle[preprint,eqsecnum,aps,epsfig]{revtex}
\tightenlines
\begin{document}
\draft
\title
{On the distribution of the total energy of a system of non-interacting
fermions: random matrix and semiclassical estimates}
\author{ O. Bohigas$^{a}$, P. Leb{\oe}uf$^{a}$ and M. J. S\'anchez$^{a,b}$}
\maketitle
\noindent
\begin{center}
{$^{a}$\it Division de Physique Th\'eorique\footnote{Unit\'e
de recherche des Universit\'es de Paris XI et Paris VI associ\'ee au CNRS.},
Institut de Physique Nucl\'eaire.\\
91406, Orsay Cedex, France.}
\center{$^{b}${\it Departamento de F\'{\i}sica, Facultad de Ciencias Exactas
 y Naturales, \\
Universidad de Buenos Aires.
Ciudad Universitaria, 1428 Buenos Aires, Argentina.}}\\ 
\end{center}
\begin{abstract} We consider a single particle spectrum as given by the
eigenvalues of the Wigner-Dyson ensembles of random matrices, and fill
consecutive single particle levels with $n$ fermions. Assuming that the
fermions are non-interacting, we show that the distribution of the total
energy is Gaussian and its variance grows as $n^2 \log n$ in the large-$n$
limit. Next to leading order corrections are also computed. Some related
quantities are discussed, in particular the nearest neighbor spacing
autocorrelation function. Canonical and grand canonical approaches are
considered and compared in detail. A semiclassical formula describing, as a
function of $n$, a non-universal behavior of the variance of the total
energy starting at a critical number of particles is also obtained. It is
illustrated with the particular case of single particle energies given by
the imaginary part of the zeros of the Riemann zeta function on the critical
line.
\end{abstract}


~~~~~~~~~~~~~~~ {\sl Submitted for publication to Physica D -- April 27, 1998.}

\pagebreak


\hfill {\sl Dedicated to our friend Boris Chirikov,}
\newline .  \hfill {\sl with respect and admiration.~~~~~~~~~~~~~~}

\vspace{1cm}

\section{Introduction}
\label{sec:int}

Consider a system of $n$ interacting fermions. The ground state total energy
may often be well approximated by macroscopic methods, a prototype of these
being the Thomas-Fermi approximation. Aside from the well-known expansion
containing the volume energy, surface energy, etc, these methods can be
further improved by the inclusion of additional systematic quantum effects,
like for example shell effects related to some symmetries of the system (see
for instance \cite{swiatecki}). Once all these contributions have been taken
into account, one would like to have some estimate of the remaining
``irreducible'' discrepancies between the calculated and measured total
energies due to non-systematic effects. These discrepancies may be viewed as
fluctuations observable as some parameters of the system are varied, like
for example the total number of particles or any other relevant external
parameter.

On the other hand, it has been established that statistical properties of
systems whose classical analog is chaotic are well described by eigenvalues
of ensembles of random matrices (see for instance \cite{bohigas}). It seems
therefore reasonable to suggest that an extreme estimate of the above
mentioned fluctuations may consist of modeling the single particle spectrum
by the spectrum of a random matrix of the Wigner-Dyson type. This may be a
crude approximation because the statistical properties of the low-lying
energy levels of the single particle sequence may not be well described by a
random matrix approach. Another shortcoming of such a model is connected to
the fact that, for simplicity, we fix to a constant the average density of
single particle states, and ignore its variations with energy. In its
simplest version, the model may apply either when only properties around the
Fermi level are considered or when particles are enclosed in a
$2$-dimensional box. However, inclusion of a variation (increase) of the
density of single particle energies can be taken care of. This model may
also be relevant in studying the energy distribution of systems of
interacting fermions trapped in a chaotic enclosure, like for example the
variations of the total energy of $n$ electrons contained in a quantum dot
as the shape of the dot is varied by some external potential or the
variation produced when varying a magnetic field.

\begin{figure}[thb]
\begin{center}
\mbox{\epsfig{file=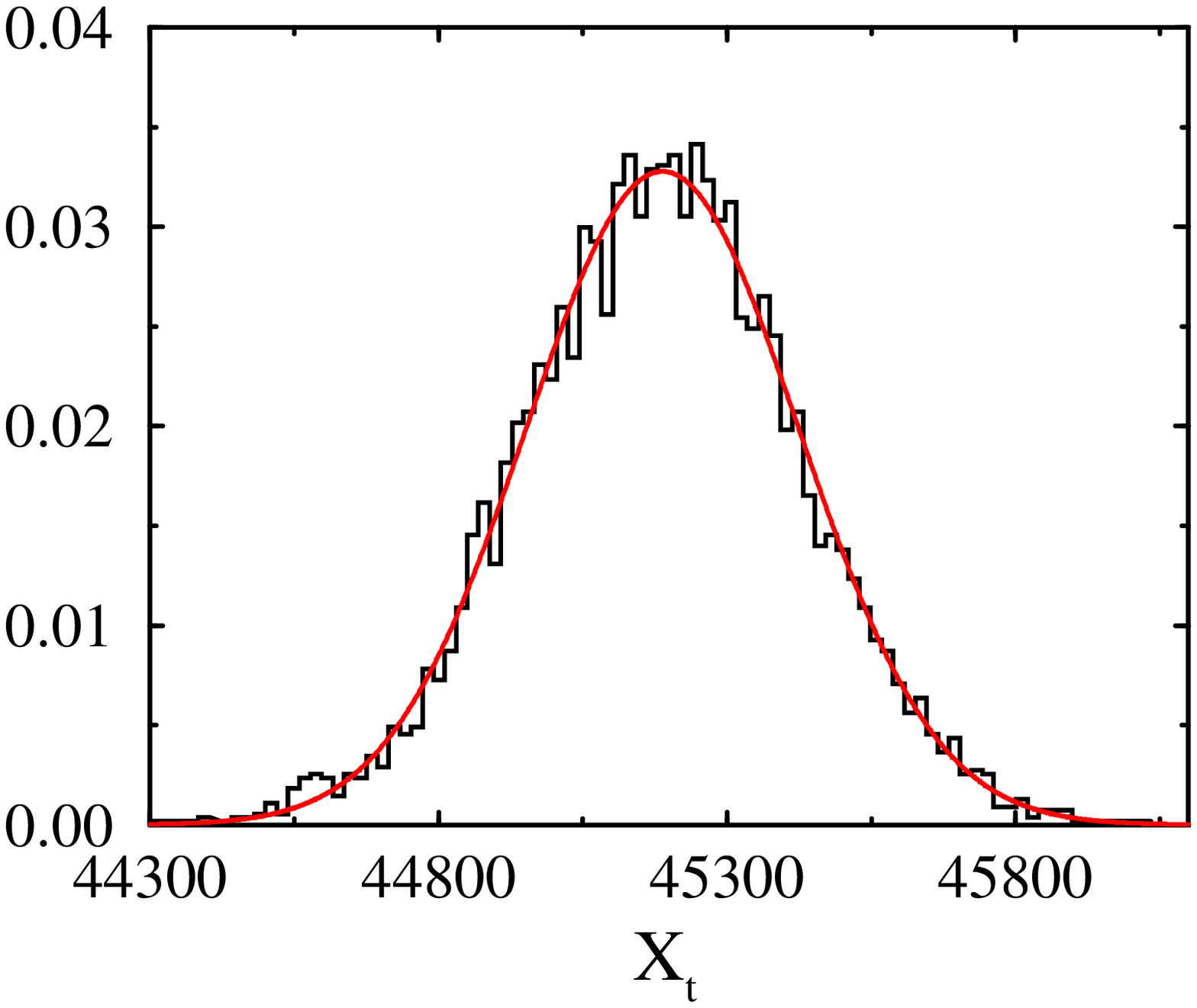,width=18cm,
bbllx=50pt, bblly=5pt, bburx=560pt,bbury=420pt,clip=}}
\end{center}
\caption{Distribution function of the total energy $X_t$ for $n=300$.
Histogram: from an ensemble of $500$ Gaussian orthogonal matrices of
dimension $N=600$; solid line: Gaussian distribution with mean value
$\langle X_{t}(300) \rangle$ and variance $\Delta^{2}(300)$ obtained from
Eqs.$(\protect{\ref{mean}})$ and $(\protect{\ref{deldac}})$, respectively.}
\label{gauss}
\end{figure}

To be specific, consider an ordered sequence of energy levels
$x_{0}<x_{1}<\ldots <x_{i}<\ldots$. Let $s_{1}, \ s_{2}, \ s_{3}, \ldots$
be the distances between consecutive eigenvalues,
$$s_i=x_i - x_{i-1} \ .
$$
We assume that the sequence is stationary and we set the mean spacing $D$ to
one. We are interested in the statistical properties of the ``ground state''
energy $X_{t}$ of a system of $n$ non-interacting fermions resulting from
adding the ``single particle'' energies $x_i$ of the $i=1,2,\ldots,n$
occupied states
\begin{equation} \label{et}
X_{t} (n)=\sum_{i=1}^n (x_{i}-x_{0})= \sum_{i=1}^n (n-i+1) \; s_{i} \; , 
\end{equation}
where $x_0$ has been taken as the energy origin. Consider now an ensemble of
single particle spectra which for the moment we let unspecified. We expect
under very general conditions the probability distribution of $X_t$ to be
Gaussian in the large $n$ limit. This, by the central limit theorem, is
obviously true for a set of uncorrelated levels. For (correlated)
eigenvalues of Gaussian ensembles of random matrices this has also shown to
be true, in fact in a more general context \cite{politzer}. As a numerical
illustration, Fig.~\ref{gauss} displays the probability distribution of
$X_t$ for $n=300$ computed for a Gaussian orthogonal ensemble of random
matrices. Our purpose is to derive expressions for the average and the
variance of $X_t$, thereby specifying completely the ``ground-state'' energy
distribution.

The first physical situation we consider is when {\sl the number} $n$ {\sl
of particles is fixed}. This case, which we call ``{\sl canonical}'' by
analogy to conventional statistical mechanics, is treated in section II.
From the mathematical point of view it is related to the autocorrelation
function of consecutive spacings, a poorly known quantity which is not a
two-point measure. Another possibility is to consider the typical
fluctuations of the total energy of the occupied levels contained in an {\sl
energy interval of given length}. In this case the number $n$ of levels
fluctuates from sample to sample around its mean value. This alternative
``{\sl grand canonical}'' problem will be treated and compared to the
canonical one in section III. The grand canonical case has the advantage of
being expressible in terms of the spectral two-point correlation function.
As expected we find that both approaches give the same asymptotic increase
of the variance of the total energy (proportional to $n^2 \log n$ in the
large-$n$ limit).

The results of sections II and III correspond to the ``universal regime'' as
given by random matrix theory. In section IV we use standard semiclassical
techniques to show that generically for ballistic -- as opposed to diffusive
\cite{diffusive} -- systems  there is a saturation effect, namely a crossover
from the $n^2 \log n$ to a $n^2$ growth of the variance of the total energy
as a function of $n$. This new regime holds when the number of particles is
larger than a system-dependent critical value and is accompanied by
non-universal oscillations around the mean growth for which we give an
explicit expression. To illustrate this point with a particular example, we
have considered the unphysical but interesting in its own and explicitly
computable case where the single particle energies are given by the
imaginary part of the zeros of the Riemann zeta function on the critical
line. Section V summarizes the results and gives some perspectives.

\section{Canonical Variance}
\label{cen}

Statistical averages over the ensemble are denoted by the symbol $\langle .
\rangle$. The mean value $\langle X_{t} \rangle$ is 
\begin{equation}
\label{mean}
\langle X_{t}\rangle=\sum_{i=1}^n (n-i+1) \langle s_{i} \rangle=
\sum_{i=1}^n (n-i+1)= \frac{n(n+1)}{2} \;,
\end{equation}
where the mean level spacing has been set equal to one.

Let us now consider the variance of $X_t$ defined, from Eq.(\ref{et}), by
\begin{eqnarray}
\Delta^{2} (n) &\equiv& \langle X_{t}^2 \rangle -\langle X_{t}\rangle^2=
\sum_{i,j=1}^n (n-i+1) (n-j+1)\; \langle s_i \; s_j \rangle -
\frac{1}{4} n^2 \; (n+1)^2  \nonumber \\ 
&=&\; \sum_{i=1}^n (n-i+1)^2  \langle s_i^2 \rangle +
\sum_{i \neq j=1}^n  (n-i+1) (n-j+1) \langle s_i s_j \rangle - \frac{1}{4} n^2 
\; (n+1)^2\; . \nonumber
\end{eqnarray}
The main quantity of interest in this expression is the autocorrelation
function $I(k)$ of spacings of consecutive levels,
$$
I(k)\equiv\langle s_{i}s_{i+k} \rangle  \;\;\; \;\;\;\;\;   \;\; \; k=1,..,n-1 \; .
$$
The stationarity of the spectrum implies that $I(k)$ depends only on the
relative index $k$ and that $\langle s_{i}^2 \rangle= \langle s^2 \rangle$,
for all $i$. Performing the sums over the index $i$
\begin{equation} \label{var}
\Delta^{2} (n)= \frac {n (n+1) (2n+1)}{6} \langle s^2 \rangle - \frac{1}{4} n^2
\; (n+1)^2  + \frac{1}{3}\sum_{j=1}^{n-1} F(n,j)\; I(j) , \; \; \;  \; \; n \ge
2 \; ,
\end{equation}
where
$$
F(n,j)\equiv j^3 - ( 3 n^2 + 3 n +1 ) j + n \ (n+1) \ (2n+1) \; .
$$
For $n=1$ the variance takes the value $\Delta^{2} (1) = \ \langle s^2
\rangle \ -1$. All the non-trivial information is contained in the
autocorrelation of spacings; a detailed study of this function will be made
in the next subsection. The extreme case of a uniform uncorrelated sequence
of levels is particularly simple since $I(j)=1$ for all $j$, and it follows
from Eq.(\ref{var}) that
\begin{equation}
\label{poisson}
\Delta^{2}_{u} = (\langle s^2 \rangle_u \ - 1) \left( \frac{n^3}{3} + 
\frac{n^2}{2} 
+ \frac{n}{6}\right) \; .
\end{equation}
When the nearest neighbor spacing distribution of the uncorrelated sequence is
$P(s)= \exp (-s)$ then ${\langle s^2 \rangle}_{u}$ in Eq.(\ref{poisson}) is
equal to $2$.

\subsection{The autocorrelation function of spacings $I(j)$}
\label{ge}
For Gaussian ensembles (GE) of random matrices, the quantities entering
Eq.(\ref{var}) can be expressed as 
\begin{eqnarray} 
\langle s^{2} \rangle  & = & 2 \;\int_{0}^{\infty} E(0,s) \; ds \; , \nonumber \\
\label{corre2}
       I(j)         & = & \;\int_{0}^{\infty} E(j,s) \; ds,\;\; \; \;\;\;\;\;  
\;j \ge 1, 
\end{eqnarray}
where $E(j,s)$ is the probability that a randomly chosen interval of length
$s$ contains exactly $j$ eigenvalues \cite{mehta}. The functions $E(j,s)$
can be written in terms of an infinite product of the eigenvalues of an
integral equation involving spheroidal functions and only numerical tables
of these functions for low values of $j$ and a limited range of $s$ exist.
Therefore the integrals in (\ref{corre2}) cannot be computed analytically
and explicit expressions for the $I(j)$ are not available. Some numerical
estimates of the integrals (\ref{corre2}), to be used in what follows, are
\cite{mehta}
\begin{equation} \label{s2}
{\langle s^{2} \rangle} = \left\{ 
\begin{array}{ll}
                         1.286, & \;\;\;\;\;\;\;\;\;\; \mbox{$\beta =1$}\\
                         1.180, & \;\;\;\;\;\;\;\;\;\; \mbox{$\beta =2$}\\
                         1.105, & \;\;\;\;\;\;\;\;\;\; \mbox{$\beta =4$}
                         \end{array}
                 \right. \nonumber
\end{equation}
where $\beta \;= 1,2$ and $4$ denote the three types of GE, orthogonal (GOE),
unitary (GUE) and symplectic (GSE) respectively. Numerical estimates
of $I(j)$ exist for low values of $j$. For instance, for $j=1$ \cite{mehta}
\begin{equation}
\label{s1s2}
I(1)= \left\{ \begin{array}{ll}
                         0.922, & \; \; \; \;\;\;\;\;\; \mbox{$\beta =1$}\\
                         0.944, &  \; \; \;\;\;\;\;\;\;  \mbox{$\beta =2$}\\
                         0.964, &   \; \; \;\;\;\;\;\;\; \mbox{$\beta =4$.}
                         \end{array}
                 \right. \nonumber
\end{equation}
For our purpose the functions $I(j)$ for arbitrary $j$ are needed and the
following ansatz
\begin{equation}
\label{ij}
I(j) = \;  1  \, - \eta /j^{2} \, - \lambda /j^4  \, - \alpha /j^{6} 
\end{equation}
will be made. The constants in Eq.(\ref{ij}) are, in principle, dependent on
the symmetry class of the GE considered ($\beta=1$, $2$, $4$). Eq.(\ref{ij})
is one of the basic equations of this paper. The simple functional
dependence on $j$ is consistent with the requirement that two sufficiently
far apart consecutive spacings should be uncorrelated. Moreover, general
considerations suggest that odd powers of $j$ may be excluded. As we
will now show, the three parameters entering Eq.(\ref{ij}) may be uniquely
determined by requiring the correct asymptotic behavior of the fluctuations
in the length of an interval containing a fixed number of levels.

Consider for that purpose the statistical fluctuations in the length $S$ of
a spacing made of $n$ consecutive nearest-neighbor spacings, $S=
\displaystyle\sum_{i=1}^{n} s_{i} \;$. The spacing variance $\sigma^2$ of $S$
can be expressed in terms of the $I(j)$ in the following manner
\begin{equation} \label{sigma2}
\sigma^2(n) \equiv \ \langle S^{2} \rangle - {\langle  S \rangle}^2 =
\; n \left( \langle s^2 \rangle - n \right)  +
2\; \sum_{j=1}^{n-1} (n-j) \; I(j), \; \;  \; \; \;\; n \geq 2 \;,
\end{equation}
while $\sigma^2(1) \equiv \langle s^{2} \rangle - 1$. We use a
non-conventional but more natural notation, namely $\sigma^2(n)$ denotes
the variance of the distribution of the sum of $n$ consecutive nearest
neighbor spacings ($n=1,2,\ldots $). The conventional notation being
$\sigma^2(n-1)$, $n=1,2,\ldots $.

Replacing the ansatz (\ref{ij}) in (\ref{sigma2}) we get
\begin{eqnarray}
\label{sigma22}
\sigma^2 (n) & = & \; \left( \langle s^2 \rangle - 1 \right) \ n \
 + 2 \ \eta \ \left[\Psi (n) +\gamma \right] + 2 \ \eta \ n \ 
 \left[ \Psi^{(1)} (n) - \zeta (2) \right] 
 + \lambda \ \left[ \Psi^{(2)} (n) + 2 \; \zeta (3) \right]  \nonumber \\ 
	 &   & + \ (n \ \lambda/3) \left[ \Psi^{(3)} (n) - 6 \ \zeta (4) \right]
	     + (\alpha/12) \left[ \Psi^{(4)} (n) + 24 \ \zeta (5) \right] 
         \nonumber \\
  &   & + \ (n \ \alpha/60) \left[ \Psi^{(5)} (n) - 120 \ \zeta (6) \right] \ ,
\end{eqnarray}
where $\gamma$ is the Euler constant, $\Psi (n) = - \gamma + \
\displaystyle \sum_{j=1}^{n-1} \frac{1}{j} $ is the DiGamma function and
$\Psi^{(m)}(z)= \frac{d^{m}}{dz} \Psi (z)$ are the PolyGamma functions
\cite{abram}. The large-$n$ behavior of $\sigma^{2}(n)$ follows from the
asymptotics of the $\Psi^{(m)}$'s \cite{abram}. Only $\Psi$ and $\Psi^{(1)}$ will be
needed, the other not contributing at the order we are working
\begin{eqnarray}
\label{asin}
\Psi(n)   & = & \log n - \frac{1}{2 n} - \frac{1}{12 n^2} + 
{\cal} O (1 / n{^4}) \;, \, \nonumber \\
\Psi^{(1)}(n)  & = &  \frac{1}{ n} + \frac{1}{2 n^2} + \frac{1}{6 n^3} + 
{\cal O} (1 / n^5) \ .
\end{eqnarray}
Then 
\begin{equation}
\label{sigmamal}
\sigma^{2}(n) = A \ n +  2 \ \eta \ \log n \ + B + \ {\cal O}(1/n^2) \ ,
\end{equation}
with 
\begin{eqnarray}
A & \equiv & {\langle s^2 \rangle} -1 - 2 \ \eta \ \zeta(2) \ 
- 2 \ \lambda \ \zeta(4) - 2 \ \alpha \ \zeta(6) \ , \nonumber \\
B & \equiv & 2 \ (\gamma + 1) \ \eta  + 2 \ \lambda \ \zeta(3) 
             + 2 \ \alpha \ \zeta(5) \ . \nonumber
\end{eqnarray}
Would one have retained a $1/j$ term in the ansatz, the leading order term
would have been of order $n \log n$.

A different quantity, closely related to the spacing variance $\sigma^2
(n)$, is the number variance $\Sigma^2 (L)$ defined as the variance of the
number of levels contained in an interval of given length $L$ taken at random.
This quantity is known analytically for the three Gaussian ensembles and has
in the large-$L$ limit the asymptotic behavior \cite{mehta}
\begin{equation} \label{sig2}
\Sigma^2 (L) = \frac{2}{\beta \pi^2} \log L + C + {\cal O} (1/L) \ ,
\end{equation}
where 
\begin{equation} \label{c}
C = 2/(\beta \pi^2) \{ \gamma + \log (2\pi) + 1 + \beta (\beta-2) \pi^2/8 +
[\log 2 - 7\pi^2/8] \delta_{\beta,4} \} \ .
\end{equation}

Consider now more carefully the relation between the spacing and the number
variances. The values that the number variance takes at integer values of
its argument are, for large $n$, related to the spacing variance through
\cite{french}
\begin{equation} \label{onesix}
\Sigma^2 (L=n) - \sigma^2 (n) = 1/6 \ .
\end{equation}
This equation indicates that: (i) as expected from their definition, the
leading order terms of both quantities coincide and, (ii) the
next-to-leading order terms differ by a constant ($1/6$), irrespective of
the symmetry class $\beta$.

Eq.(\ref{onesix}) completely determines all the unknown coefficients in
Eq.(\ref{ij}). In fact, consistency with Eq.(\ref{onesix}) imposes the
following conditions in Eq.(\ref{sigmamal})
\begin{eqnarray}
\label{parametros}
 \eta            & =   & \frac{1}{\beta \pi^{2}} \ , \nonumber \\
 A         & =   & 0 \ ,   \\ 
 C - B   & =   & 1/6 \ .   \nonumber 
\end{eqnarray}
The remaining free parameters $\lambda$ and $\alpha$ are adjusted to satisfy
the second and third conditions
\begin{eqnarray}
\label{alfas}
\lambda & = & \frac{\langle s^2 \rangle - 1 - \ 2 \ \eta \ \zeta(2)}
{2 \ \zeta(4)} - \frac{\zeta(6)}{\zeta(4)} \alpha \ , \nonumber \\
\alpha & = & \frac{\zeta(4)}{2 [\zeta(3) \zeta(6) - \zeta(4) \zeta(5)]}
\left\{2 \ \eta \ (\gamma + 1) - C + \frac{1}{6} 
+ \ \frac{\zeta(3)}{\zeta(4)} 
\left[ \langle s^2 \rangle - 1 - 2 \ \eta \ \zeta(2) \right] \right\} \ .
\end{eqnarray}
Using Eq.(\ref{s2}) these two constants take the values $\lambda = 0.02027,
 \ 0.03013, \ 0.00997$ and $\alpha = - 0.04483, \ - 0.02550, \ 4\times
 10^{-5}$ for $\beta=1,2$ and $4$, respectively. 

\begin{figure}[thb]
\begin{center}
\mbox{\epsfig{file=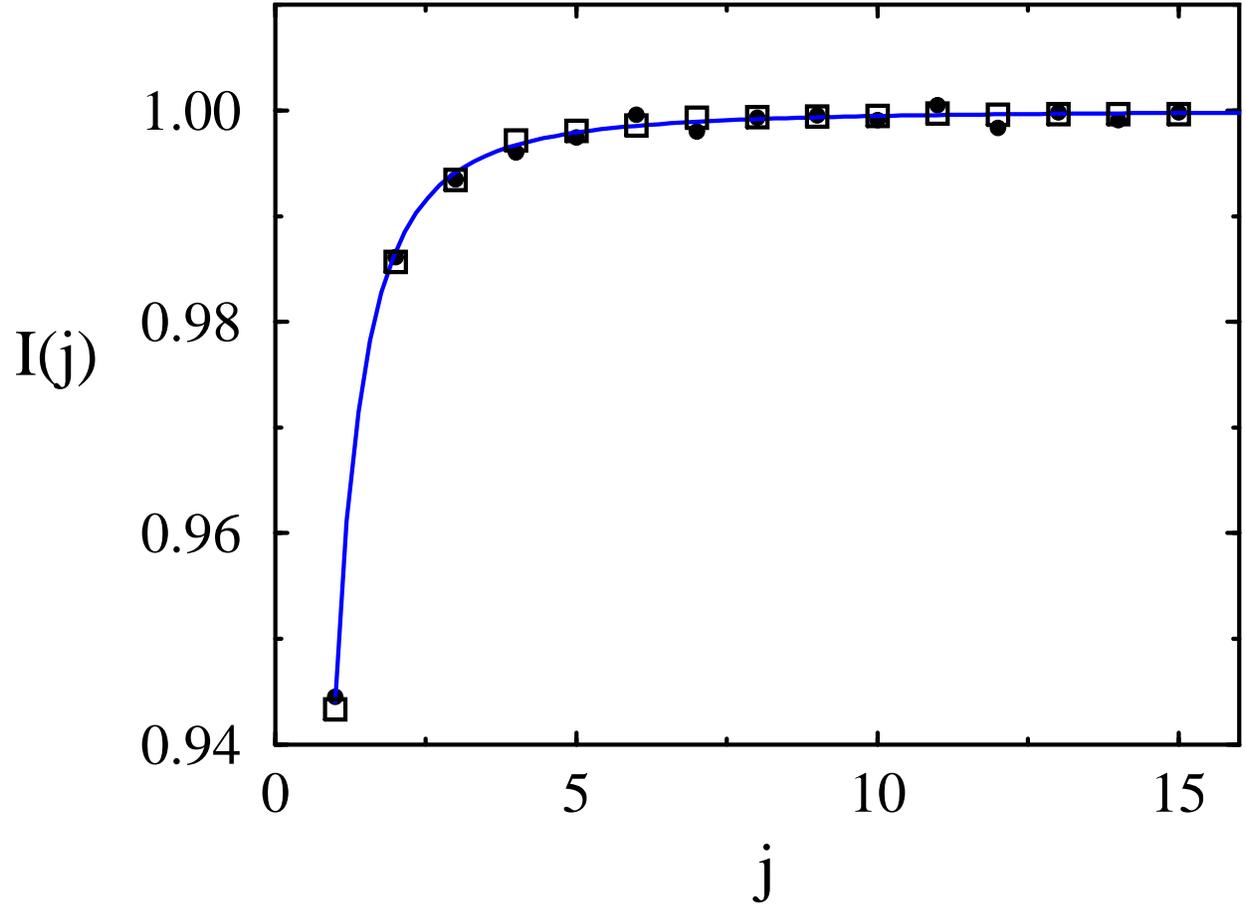,width=17cm,
bbllx=-10pt, bblly=-20pt, bburx=526pt,bbury=415pt,clip=}}
\end{center}
\caption{Autocorrelation of spacings $I(j)$ for $\beta=2$. Dots: obtained from an ensemble of 
$500$ Gaussian unitary matrices of dimension $N=500$; solid line:
the ansatz $(\protect{\ref{ij}})$ for $\beta = 2$; squares:  computed 
from the zeros of the Riemann zeta function (taken from [9]).
}
\label{sisjgue}
\end{figure}

In order to test the accuracy of Eq.(\ref{ij}) we have computed numerically
the autocorrelation function $I(j)$ for orthogonal as well as unitary random
matrices. As can be seen from Fig.~\ref{sisjgue} the agreement between the
ansatz and the numerical simulations for $\beta=2$ is very good. Because the
overall agreement is of similar quality we do not show the analogous curve
for GOE. As an additional quantitative test let us mention that when
expression (\ref{ij}) is evaluated at $j=1$ it reproduces the values
in Eq.(\ref{s1s2}) with a $1 \times 10^{-3}$ error for any $\beta$. This is
remarkable if one keeps in mind that only asymptotic information has been
used when determining the parameters in the ansatz (\ref{ij}).

In a previous study of the autocorrelation function of the spacings Odlyzko
\cite{odlyzko} has quoted a conjecture proposed by Dyson: $I(j) \approx
1 - \eta/ j^{2}$ (see also chapter 16 in \cite{mehta}). This ansatz
coincides to lowest order with Eq.(\ref{ij}). Aside from considerations
related to accuracy, let us emphasize that the inclusion of a term of higher
order (like $ 1/ j^4$ or $1/j^6$) in (\ref{ij}) is essential in order to
obtain the correct asymptotic behavior for $\sigma^{2}(n)$ as well as, as we
shall see later, for $\Delta^{2}(n)$. This is so because these higher order
terms ensure the vanishing of the linear term in Eq.(\ref{sigmamal}).

Eq.(\ref{ij}) is consistent with known relationships satisfied by the
autocorrelation function. For example, the sum rule
$$
2 \sum_{j=1}^\infty \left[I(j) - 1 \right] + \sigma^2 (1) = 0
$$
valid for GE of random matrices \cite{pandey} is equivalent to the condition
$A=0$ in Eq.(\ref{parametros}) (i.e., the vanishing of the linear term of
$\sigma^2 (n)$ \cite{bhp}).

\subsection{Variance of the total energy}

We have now the necessary ingredients to compute the variance of the
total energy. Replacing in Eq.(\ref{var}) the ansatz (\ref{ij}) and
using the definition of the DiGamma and PolyGamma functions it follows that
\begin{eqnarray} 
\Delta^{2}(n) & = &  
\frac{n^2}{2}  \left( C -\frac{7\eta}{3} - \frac{1}{6} \right) +
 \frac{n}{2} \left( C -\frac{5\eta}{3} - \frac{1}{6} \right) + 
 \frac{1}{6} \left( C - 2\eta  - 2 \ \gamma \ \lambda 
 - 2 \ \alpha \ \zeta(3) - \frac{1}{6} \right) \nonumber \\ 
& & + \Psi(n) \left[ \eta \left(n^2+n+\frac{1}{3}\right) -
\frac{\lambda}{3} \right] +
\Psi^{(1)} (n) \ \eta \left( \frac{2 n^3}{3}  + n^2 + \frac{n}{3} \right) 
\nonumber \\
& & + \frac{1}{2} \Psi^{(2)} (n) \left[ \lambda \left( n^2 + n + \frac{1}{3} 
\right) - \frac{\alpha}{3} \right] + \frac{1}{6} \Psi^{(3)} (n) \ \lambda \left( 
\frac{2 n^3}{3}+ n^2 + \frac{n}{3} \right) \nonumber \\
& & + \frac{1}{24} \Psi^{(4)} (n) \ \alpha \left( 
n^2 + n + \frac{1}{3} \right) + \frac{1}{120} \Psi^{(5)} (n) \ \alpha \left( 
\frac{2 n^3}{3}+ n^2 + \frac{n}{3} \right) \nonumber 
\end{eqnarray}
for $n\ge 2$. Using again the asymptotic expressions (\ref{asin}) for 
$\Psi (n)$ and $\Psi^{(1)} (n)$, as well as those of $\Psi^{(2)} (n)$ and 
$\Psi^{(3)} (n)$ \cite{abram} we finally get
\begin{eqnarray} \label{deldac}
\Delta^{2}(n)      & = &  \eta \; n^2 \log n + \frac{1}{2} \left(
C -
\eta - \frac{1}{6} \right) n^2 + \eta \;  n \log n + 
\frac{1}{2} \left( C - \frac{1}{6} \right) n 
\nonumber \\ 
&   & + \frac{1}{3} \; (\eta - \lambda) \log n  \ +  
\frac{1}{6} \left[ C + \frac{(\eta -1)}{6}  - 
2 \left( \gamma + \frac{5}{6} \right) \lambda - 2 \ \alpha \ \zeta(3) \right] + 
{\cal O} (1 / n)    \; . 
\end{eqnarray}
Notice that, as it happened for $\sigma^2 (n)$ with the terms of order $n
\log n$ and $n$, the functional form of the ansatz (\ref{ij}) together with
the method by which the parameters $\eta$, $\alpha$ and $\lambda$ have been
determined insure the vanishing of the terms of order $n^3 \log n $ and
$n^3$ of the variance of the total energy. That the remaining leading order
behavior $n^2 \log n $ in Eq.(\ref{deldac}) is correct is further confirmed
by the fact that it coincides with the asymptotic result obtained from the
linear statistic theory (see section III). This behavior is in contrast with
the faster growth of order $n^3$ of an uncorrelated spectrum (cf
Eq.(\ref{poisson})). The difference is of course due to the rigid nature of
the GE spectra.

\begin{figure}[thb]
\begin{center}
\mbox{\epsfig{file=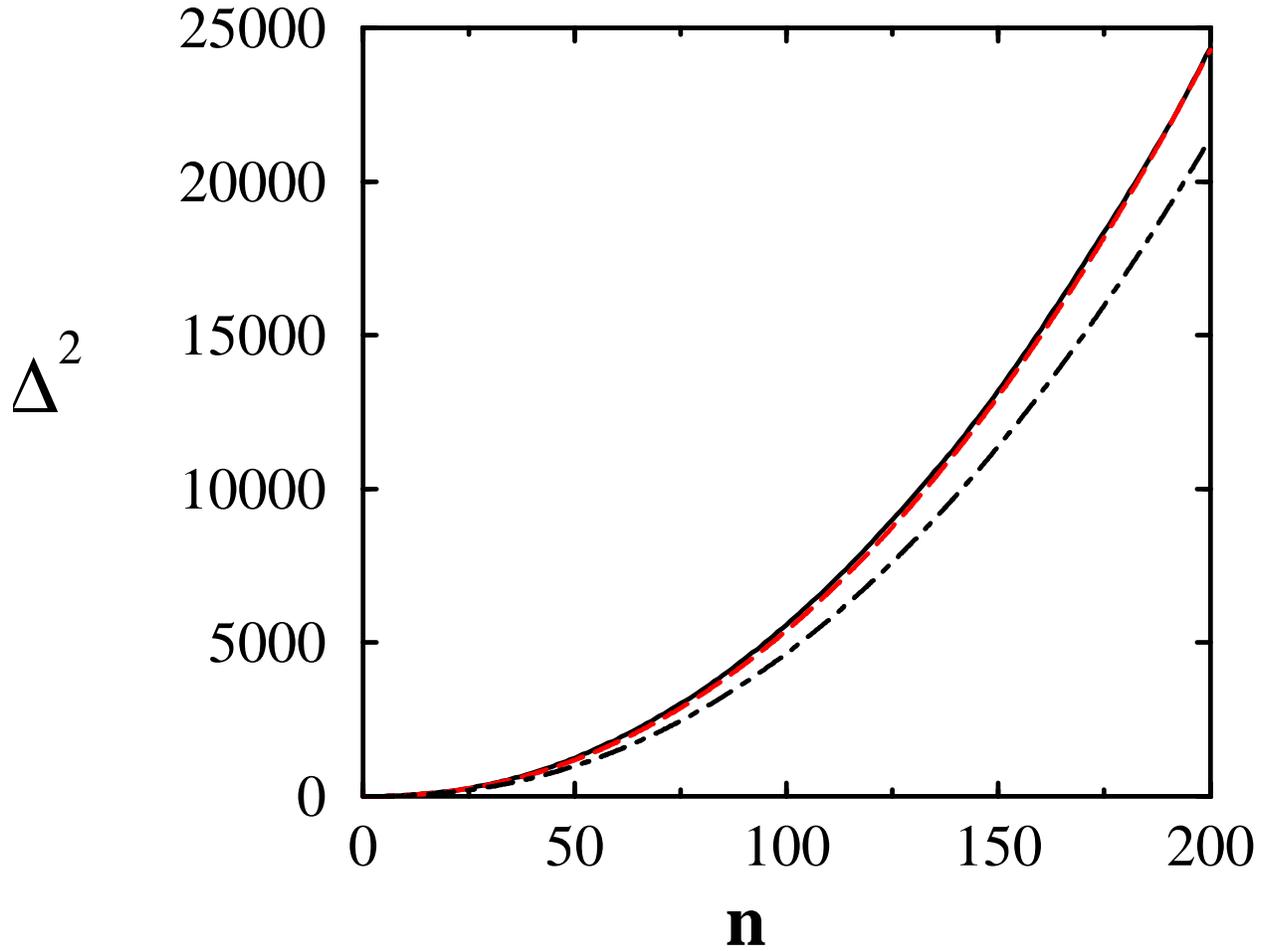,width=17cm,
bbllx=32pt, bblly=30pt, bburx=511pt,bbury=443pt,clip=}}
\end{center}
\caption{Canonical number variance $\Delta^{2}$ for $\beta =1$ as a function of
$n$. Solid line: from an ensemble of $500$ Gaussian orthogonal matrices of
dimension $N=600$; dashed line: theory (see Eq.$(\protect{\ref{deldac}})$); 
dot-dashed line: leading order term of the same equation.}
\label{grandegoe}
\end{figure}

Fig.~\ref{grandegoe} shows for GOE the comparison of numerical results with
Eq.(\ref{deldac}) for the variance of the total energy as a function of $n$.
The overall agreement is good (for $n \approx 50$ the relative
error is less than $ 2 \% $).  The theoretical curve is sensitive to
the precise value of $\langle s^{2} \rangle$ which enters in the definition
of $\lambda$ and $\alpha$, and is known only up to the third digit (cf
Eq.(\ref{s2})). For comparison we have also plotted in Fig.~\ref{grandegoe}
the leading-order term of $\Delta^2$, which clearly fails to reproduce
accurately the numerical results in the interval of $n$ displayed (the error
is of the order of 10\% at $n\approx 50$).

In the previous analysis we have taken a single particle level ($x_0$ in
(\ref{et})) as the reference energy. In some physical applications it may be
more appropriate to measure energies with respect to an arbitrary origin,
which will not coincide with a single particle level but rather will lie in
a random position between two of them. As shown in Appendix A when this
construction is adopted, namely filling $n$ successive levels located just
above this origin, Eq.(\ref{mean}) giving the mean value of the energy is
modified as follows $n (n+1)/2 \rightarrow n^2 /2$, whereas for the variance
of the energy one has
\begin{equation} \label{origin}
\Delta^2 (n) \rightarrow \Delta^2 (n-1) + \frac{n^2}{2} \left( \frac{1}{2} -
\frac{\langle s^2 \rangle}{3} \right) + \frac{n}{2} \sigma^2 (n) \ .
\end{equation}
Notice that both sides of (\ref{origin}) have the same leading order, though
different coefficients for higher order terms.

\section{Grand canonical Variance}

Here we treat a slightly different problem with respect to the previous
section. Instead of fixing the number of occupied levels, we consider an
energy interval of given length $2L$ and compute the energy variance for the
single particle energies $x_1,\ldots,x_n$ contained in it (to follow
conventional notation, in this section we will denote the length of the
energy interval by $2 L$, whereas in the previous section we were denoting
it by $L$). The length $2L$ of the energy interval is now kept fixed but the
number $n$ of levels contained in it may now vary from sample to sample.
This is a particular example of a more general class of problems usually
called ``linear statistic'' in random matrix theory \cite{dyson}, dealing with
the distribution of a variable $W$ of the form
\begin{equation}\label{00}
W = \sum_{i=1}^n f(x_{i}) \; ,
\end{equation}
where $f$ is an arbitrary function. The variance of $W$ was given in
\cite{dyson}
\begin{equation}\label{0}
V_W = \int\int_{-L/D}^{L/D} d x \ d y \left[ \delta(x-y)-Y_2 (x,y)\right] f(xD) 
f(yD)
\end{equation}
or, alternatively, in the Fourier space
\begin{equation} \label{1}
V_W = \frac{1}{D}\int_{-\infty}^\infty d k\left[1-b(D k)\right] \varphi(k)
\varphi(-k) \ .
\end{equation}
Here $D$ is the mean level spacing (assumed to be constant; it will
be set to one at the end), $Y_2$ the two-level cluster function,
$b(k)=\int_{-\infty}^{\infty} Y_{2}(x) \; \exp[2 i
\pi k x] \; dx $ the form factor and $\varphi (k)$ the Fourier transform of
$f(x)$ 
$$
\varphi(k)= \int_{-L}^{L} f(x) \; \exp[-2 i \pi k x] \; dx
$$
($f(x)$ is assumed to be zero outside the interval $[-L,L]$).

To compute the number variance via Eq.(\ref{1}), one must choose  $f(x)=1$,
as Dyson and Mehta did. For our purpose, namely to compute the variance of
the total energy, $f(x)$ should be chosen as follows
\begin{equation} \label{1p}
f(x) = \left\{ \begin{array}{ll}
     x + L \;\;\;\;\;\;\;  & \mbox{$-L\leq x \leq L$} \\
     0 \;\;\;\;\;\;\;\;\;\;\;  & \mbox{$|x|>L$}
		\end{array} 
		\right. \nonumber  
\end{equation}
whose transform is 
\begin{equation} \label{2}
\varphi (k)=\frac{\rm i}{2 \pi^2 k^2} [2\pi L k \cos (2\pi L k)-\sin(2\pi L k)]
+ \frac{L}{\pi k}\sin(2\pi L k) \ . 
\end{equation}
In Eq.(\ref{1p}) we have chosen to measure the energy with respect to the
lower (or equivalently upper) end of the energy interval considered. This,
which is somewhat natural and analogous to Eq.(\ref{et}), is by no means
compulsory and the reference may be taken at any given point of the energy
axis. The variance of the total energy depends on the location of the
reference point, the minimum being obtained when it is at the center of the
interval. A general relation connecting the variance at different reference
points is given in Appendix B.

Let us now consider the case of a large (average) number $n$ of levels
contained in the interval. Though we are using for notational simplicity the
same symbol as in the canonical case, for grand canonical expressions $n$
will mean the expectation value $\langle n \rangle =2L/D$. Therefore, in
contrast to the canonical case, it takes continuous values. Following
Ref.\cite{dyson}, we split the integral in Eq.(\ref{1}) in two parts
\begin{equation} \label{4}
V_{E}= \frac{2}{D} \int_0^\epsilon \varphi (k) \varphi (-k) \ [1-b(D k)] \ d k +
\frac{2}{D} \int_\epsilon^\infty \varphi (k) \varphi (-k) \ [1-b(D k)] \ d k
\end{equation}
where we have used the parity of the integrand. The parameter $\epsilon$ is chosen
such that $1/L\ll \epsilon \ll 1/D$. For $k<\epsilon$, $D k$ is always much smaller than
one and the form factor can be approximated by \cite{mehta} 
$$
b(D k) \approx 1 - \frac{2 D}{\beta} \ |k|
$$
and the first term in the r.h.s. of Eq.(\ref{4}) (denoted $V_{E}^{(a)}$) can
be written 
$$
V_{E}^{(a)} = \frac{4 L^2}{\beta \pi^2} \int_0^{2\pi L\epsilon} \ \left[ 
\frac{1}{x} - \frac{2 \cos x\sin x}{x^2} + \frac{\sin^2 x}{x^3} \right] d x \ ,
$$
which when integrated and taken in the limit $L\epsilon \gg 1$ gives
\begin{equation}\label{5}
V_{E}^{(a)} \approx \frac{4 L^2}{\beta \pi^2} \, [\log(2\pi L\epsilon)+\gamma
+\log 2  -1/2] \ . 
\end{equation}
On the other hand, for $k>\epsilon$ we have $k L\gg 1$ and the function
$\varphi (k)$ shows rapid oscillations as compared to the variations of the
form factor. We then replace the product $\varphi(k) \varphi(-k)$ in the
second term of the r.h.s. of Eq.(\ref{4}) (denoted $V_{E}^{(b)}$) by its
average value over $k$
$$
\langle \varphi (k) \varphi (-k) \rangle_k = \frac{L^2}{\pi^2 k^2} + 
\frac{1}{8 \pi^4 k^4}
$$
and then
$$
V_{E}^{(b)} \approx - \frac{2 L^2}{\pi^2} \int_{D \epsilon}^\infty \frac{[b(x)-1]}
{x^2} \  d x
- \frac{D^2}{4 \pi^4} \int_{D \epsilon}^\infty \frac{[b(x)-1]}{x^4} \ d x \ .
$$
The second integral in the latter equation is of order $1/(\epsilon L)^2$ with
respect to the first one and therefore can be neglected. Evaluating the first
integral we get
\begin{equation} \label{5p}
V_{E}^{(b)} \approx \frac{-4 L^2}{\beta
\pi^2} \left[\log(D\epsilon)-1\right] + (\beta-2) \frac{L^2}{2} \ . 
\end{equation}
Collecting the two terms we obtain the final expression for the variance
which we express in terms of the {\sl average number} $n$ of levels
contained in the interval and the constant $C$ (cf (\ref{c})) (we
put moreover $D=1$)
\begin{equation} \label{6}
V_{E}(n) \approx \eta \ n^2 \log n + \frac{1}{2} 
\left( C - \eta \right) n^2  \ .
\end{equation}

Let us now compare grand canonical and canonical results. When the sequence
considered is large ($n\gg 1$), we expect the energy variance computed
for a fixed number $n$ of occupied levels ($\Delta^2$, canonical) to be
equal to the energy variance when the fixed length of the energy interval
contains on the average $n$ levels ($V_{E}$, grand canonical). The
comparison of Eqs.(\ref{6}) and (\ref{deldac}) shows that indeed to leading
order $V_{E} = \Delta^2$ (the same leading order behavior has been obtained
in Ref.\cite{wilkinson} for the particular case of GOE employing Dyson
Brownian motion). However the correction terms are different. It is
interesting to note that the next to leading order corrections differ, as in
Eq.(\ref{onesix}) for the number and spacing variances, by a constant
independent of $\beta$
\begin{equation}\label{6p}
\lim_{n\rightarrow \infty} \frac{V_E (n) - \Delta^2 (n)}{n^2} = \frac{1}{12}
\ ,
\end{equation}
which follows from Eqs.(\ref{deldac}) and (\ref{6}).

To conclude this subsection let us point out that exact expressions for the
energy variance can be computed in the grand canonical approach for any
value of $n$ without assuming that it is large. For example for the unitary
case $(\beta=2)$ from the definition Eq.(\ref{1}), using Eq.(\ref{2}) and
the fact that
\begin{equation} \label{7}
b(k)= \left\{ \begin{array}{ll}
      1 - \; \mid k \mid  \;\;\;\;\;\;\; & \mbox{$\mid k \mid\leq 1$} \\
   0  & \mbox{$\mid k \mid > 1$}
		\end{array} 
		\right. \nonumber  
\end{equation}
it follows that
$$
V_{E} (n)= \frac{n^2}{2 \pi^2} \left[ \begin{array}{l} \displaystyle \log n
- {\rm Ci} (2 \pi n) - \frac{4\pi n}{3}  {\rm Si} (2 \pi n) + \frac{\sin (2\pi
n)}{6\pi n} + \left(\frac{1}{12\pi^2 n^2} - \frac{2}{3} \right) \cos (2\pi n)
\end{array} \right. 
$$
\begin{equation} \label{8}
\left. \begin{array}{l} - \displaystyle \frac{1}{12\pi^2 n^2} + \frac{2\pi^2
n}{3} + \gamma +\log(2\pi) + \frac{1}{2} \end{array} \right] \ , \;\;\;\;\;\;\;\; \beta=2 \ ,
\nonumber 
\end{equation}
where ${\rm Ci} (x)$ and ${\rm Si} (x)$ are the CosIntegral and SinIntegral
functions\cite{abram}, respectively. $V_E (n)$ grows like $n^3$ for
$n\rightarrow 0$ (remember that here $n$ is a continuous variable) and then
switches to the $n^2 \log n$ behavior for larger values of $n$.

\section{Semiclassical treatment: Non-universalities}

\subsection{Chaotic Systems}

The strength and vitality of random matrix theories rely heavily on the
universality of some of its predictions. As a counterpart it is poorly
adapted to capture some system-specific features. However special tools have
been developed for that purpose to study systems whose classical analog is
chaotic. Indeed, one of the important achievements of semiclassical theories
has been to determine the limits of validity of the universal regime as
described by random matrix theories by including system-dependent
corrections \cite{berry}. Our purpose now is to adapt these methods, which
have been developed for the form factor and are thus applicable to the
grand canonical case, to the study of the variance of the total energy.

Our starting point will be Eq.(\ref{1}) written in a slightly different form
\begin{equation} \label{10}
V_W = \frac{2}{D^2}\int_{0}^\infty d \tau K(\tau)
\varphi(\tau/D) \varphi(-\tau/D) \ .
\end{equation}
Here $K(\tau) = 1-b(\tau)$ and we have used the parity of the integrand. For
simplicity we restrict in the following to systems having no time reversal
symmetry ($\beta=2$). The results can be easily generalized to the other
symmetry classes.

In the previous section we computed the universal behavior of the variance
of the total energy by inserting in (\ref{10}) the random matrix spectral
form factor (cf Eq.(\ref{7}) for the unitary ensemble). It is possible
however to give a more accurate description of the form factor based on
semiclassical approximations. For systems having a classical analog the key
ingredient is a formula expressing the spectral density $\rho (E) =
\displaystyle \sum_j \delta (E-E_j)$ as a sum over the classical 
periodic orbits
\begin{equation} \label{10p}
\rho(E) \approx \langle \rho(E)\rangle_E + \frac{1}{\pi\hbar}\sum_p
\sum_{r=1}^\infty A_{p,r} \cos\left(r S_p / \hbar\right) \ .
\end{equation}
Here $\langle \rho(E)\rangle_E=1/D$ is the spectral average density of states, 
$A_{p,r} = T_p/\sqrt{|\det(M_{p}^r - 1)|}$ is an amplitude that depends on the 
period $T_p$ and the stability matrix $M_p$ of the primitive orbit $p$, $r$ 
are the repetitions and $S_p$ is the action \cite{gutz}. We restrict our
considerations to chaotic systems for which periodic orbits are isolated
and unstable.

Using Eq.(\ref{10p}) for the density of states Berry \cite{berry}
proposed the following form for $K(\tau)$ for chaotic systems
\begin{equation} \label{11}
K(\tau)= \left\{ \begin{array}{ll}
  (D/h) \displaystyle \sum_{p,r} A_{p,r}^2 \ \delta (T- r T_p) \;\;\;\;\;\;\;\; & 
  \mbox{$\tau < \tau^* $} \\
  \tau  \;\;\;\;\;\;\; & \mbox{$ \tau^* < \tau \leq 1$} \\
   1  & \mbox{$\tau > 1$} \ ,
		\end{array} 
		\right. \nonumber  
\end{equation}
where $h$ is Planck's constant and $\tau=T D/h$ is the (rescaled) time
measured in units of the Heisenberg time. The main physical ingredient
entering this formula is a classical sum rule due to Hannay and Ozorio de
Almeida \cite{ho} which takes into account the exponential proliferation of
long periodic orbits as well as their ergodicity. As a consequence of this
as well as other semiclassical considerations the form factor for times
$\tau$ larger than a certain critical time $\tau^*$ becomes ``universal''
(i.e., coincides with random matrix theory). The situation is different for
short periodic orbits which do not display this universality and whose
contributions, explicitly written down in Eq.(\ref{11}), produce
system-dependent deviations from random matrix theory for $\tau<\tau^*$.
Usually the parameter $\tau^*$ is chosen in order to satisfy $\tau_{\rm min}
\ll \tau^* \ll 1$, where $\tau_{\rm min}=T_{\rm min} D/h$ is the rescaled
period of the shortest periodic orbit.

The contribution of short periodic orbits to the variance of the total
energy is obtained by replacing $K(\tau)$ for $\tau<\tau^*$ in
Eq.(\ref{10}), with the result
\begin{equation}\label{11p}
\frac{2}{h^2} \sum_{p,r} {\hat A}_{p,r} \ \varphi(r \tau_p/D) \ \varphi(-r
\tau_p/D) \ .
\end{equation}
Here ${\hat A}_{p,r}=h^2 \tau_p^2/(D^2 \sqrt{|\det(M_{p}^r - 1)|})$ is the
usual amplitude $A_{p,r}$ written in terms of the rescaled time $\tau_p$.

For times $\tau>\tau^*$ we can repeat the same steps as in the previous
section (the form factor coincides with GUE), with the important difference
however that now there is a cutoff in the integral (\ref{10}) at
$\tau=\tau^*$. The variance of the total energy may now be written in the
following form
\begin{equation} \label{12}
V_E (n) = V_E^{(1)} (n) +  V_E^{(2)}(n) +  V_E^{(3)}(n) \ ,
\end{equation}
where $V_E^{(1)}$ is the GUE result (\ref{8}) and
\begin{eqnarray}\label{12p}
V_{E}^{(2)}(n) & = & - \frac{D^2 n^2}{2 \pi^2} \left[\log(n\tau^*)+
  \frac{\cos(2\pi n\tau^*)}{(2\pi n\tau^*)^2}+\frac{\sin(2\pi n\tau^*)}{2\pi n\tau^*} -
  {\rm Ci} (2 \pi n\tau^*) - \frac{1}{(2 \pi n\tau^*)^2} \right. \nonumber \\
&   & \left. \;\;\;\;\;\;\;\;\;\;\;\;\;  + \log(2\pi) + \gamma - \frac{1}{2} \right] \ , \nonumber \\
 V_{E}^{(3)}(n) & = & \frac{D^4 n^2}{2\pi^2 h^2} \sum_{r\tau_p<\tau^*} \frac{{\hat
    A}_{p,r}^2}{r^2 \tau_p^2} \left\{\left[\cos(\pi n r\tau_p) - \frac{\sin
  (\pi n r\tau_p)}{\pi n r\tau_p}\right]^2 + \sin^2 (\pi n r\tau_p)\right\} \ .
\end{eqnarray}
$V_{E}^{(2)}$ is the contribution to the variance due to the lower limit at
$\tau=\tau^*$ in the integral (\ref{10}). Aside from the system-dependent
parameter $\tau^*$, its functional form is general. $V_{E}^{(3)}$ is
obtained from (\ref{11p}) using the explicit form (\ref{2}) of the Fourier
transform $\varphi$ and contains detailed system-specific information.

The energy variance as a function of the number of particles exhibits now
two different regimes. For $n\ll 1/\tau^*$, $V_E^{(2)}$ and $V_E^{(3)}$ are
of order $(n\tau^*)^2$ and can therefore be neglected. The energy variance
is then given by the random matrix expression. For $n \approx 1/\tau^*$
these two terms can no longer be neglected and there is a crossover to a
different regime, not described by random matrix theory. This is clearly
seen for $n\gg 1/\tau^*$ when there is an almost perfect cancellation
between $V_E^{(1)}$ and $V_E^{(2)}$, leading to
\begin{equation}\label{12b}
 V_E (n) \approx \frac{D^2 n^2}{2 \pi^2} \left[ 1-\log\tau^* + {\cal O}
 (1/n\tau^*)^2 \right] + V_E^{(3)} (n) \;\;\;\;\;\;\;\;\;\;\;\;\;\; 
 n \gg  1/\tau^* \ .
\end{equation}
From Eq.(\ref{12p}) for $V_E^{(3)}$ we then see that now the variance
instead of growing as $n^2 \log n$ like in random matrix theory, saturates
and increases as $\kappa \ n^2$. The prefactor $\kappa$ is not constant but
shows non-universal oscillations as a function of $n$ described by the
short-periodic-orbit contributions. In order to determine its average value
over $n$ we replace each oscillating term between curly brackets in
Eq.(\ref{12p}) by its mean value (which is equal to one). Then, denoting
this average by an upper bar
$$
\overline{V_{E}^{(3)}} = \frac{D^4 n^2}{2\pi^2 h^2} \sum_{r\tau_p<\tau^*} 
\frac{{\hat A}_{p,r}^2}{r^2 \tau_p^2}
= \frac{D^2 n^2}{2\pi^2} \sum_{p,r} \frac{1}{r^2 |\det(M_{p}^r - 1)|} \ ,
$$
where the last sum extends over all orbits satisfying $r T_p < h \tau^* /D$.
An estimate of $\overline{V_E^{(3)}}$ may be obtained by replacing this sum
by an integral weighted by the density $\exp(h_{KS} T)/T$ over the period
$T$ of the periodic orbits ($h_{KS}$ is the Kolmogorov-Sinai entropy)
\cite{gutz}. This approximation, which is not well justified in this regime
of ``short periodic orbits'', will be shown however to give good results for
the zeros of the Riemann zeta function (see next subsection). Then using the
Hannay-Ozorio de Almeida sum rule $|\det(M_{p}^r - 1)|
\rightarrow \exp(h_{KS} T)$ one has
\begin{equation}\label{12bb}
\overline{V_{E}^{(3)}} \approx \frac{D^2 n^2}{2\pi^2} \int_{T_{\rm min}}^{h
\tau^*/D} \frac{dT}{T} = \frac{D^2 n^2}{2\pi^2} \log(\tau^*/\tau_{\rm min}) \ .
\end{equation}
Inserting (\ref{12bb}) in (\ref{12b}) we finally obtain 
\begin{equation}\label{12c}
 \overline{V_E (n)} = \frac{D^2 n^2}{2 \pi^2} \left[ 1-\log\tau_{\rm min} + {\cal O}
 (1/n\tau^*)^2 \right] \ .
\end{equation}
Thus in this approximation the average of the energy variance in the
non-universal regime is determined by a {\sl single parameter}, namely the
rescaled period of the shortest periodic orbit.

Not only is the average of the energy variance determined by the shortest
period. For $n\gg 1/\tau_{\rm min}$ the contribution of the term $\sin(\pi n
r \tau_p)/(\pi n r \tau_p)$ in Eq.(\ref{12p}) is negligible for any
$\tau_p$. Then, in this limit, each oscillating term within curly brackets
is again equal to one and therefore the energy variance itself is given by
the r.h.s. of Eq.(\ref{12c}). In summary, in the non-universal regime the
energy variance shows, superimposed to an $n^2$ growth, oscillations whose
amplitude is damped with increasing $n$. As the number of contributing terms
in (\ref{12p}) diminishes, the structure of the oscillations becomes more
regular and eventually only the lower frequencies survive just before the
complete extinction of the oscillations.

It is worth noticing also that the damping of the oscillations is a
remarkable peculiarity occurring only when the reference point used to
measure the single particle energies is at one border of the energy interval
considered. For example, setting the origin at the center of the interval,
no damping is found and the amplitude of the fluctuations remains constant
as a function of $n$. This latter behavior with non-universal oscillations
of constant amplitude is similar to what happens for the other two-point
measures mainly considered so far, namely the number variance and the
Dyson-Mehta least square statistic \cite{berry,berry2}.

It would be valuable to test the previous results (and in particular
Eq.(\ref{12})) as a function of $n$ for a real physical system, like for
example by locating of the order of $100$ particles inside a billiard of
nanometric size as in the coulomb blockade experiments in quantum dots
\cite{qdots}. Another possibility, not related to experiments but which can
be explicitly computed, is the Riemann zeta function.

\subsection{Application to the Riemann zeta function}

This function constitutes a paradigm in the field of quantum chaos. There
are many reasons for that. On the one hand the density of zeros of $\zeta
(1/2 + {\rm i} E)$ as a function of $E$ may be expressed through the
equation
\begin{equation}\label{13}
\rho_{\zeta}(E) = \langle \rho_\zeta (E)\rangle_E - \frac{1}{\pi}\sum_p
\sum_{r=1}^\infty \frac{\log p}{p^{r/2}} \cos\left(r E \log p \right) \ .
\end{equation}
The sum here goes over all the prime numbers $p=2,3,5,\ldots$ and $\langle
\rho_\zeta (E)\rangle_E = 1/D = (1/2\pi) \log(E/2\pi)$ (we are systematically
ignoring here problems related to the convergence of the series).
Eq.(\ref{13}) has the same structure as Eq.(\ref{10p}) and this suggests a
dynamical interpretation of it. On the other hand, it has been shown
numerically \cite{odlyzko} and proved in some cases and demonstrated by
heuristic arguments in others \cite{gm,odlyzko,berry,bk} that many
statistical properties of the critical zeros of that function coincide with
the GUE results of random matrix theory.

Because of the resemblance of the properties of the critical zeros of the
Riemann zeta function with those of the eigenvalues of a real chaotic
system, and because of the possibility of performing explicit computations
for that function, our purpose now is to use the imaginary part of the zeros
as the ``single particle spectrum'' of some unknown physical system. As a
first test of the results obtained in this paper we have considered the
nearest neighbor spacing autocorrelation function $I(j)$ for a set of $50\
000$ zeros located around the $10^{12}$th zero of $\zeta (1/2+{\rm i}E)$
(the set starts at $E_0 = 267653395648.8475\ldots$) computed by Odlyzko
\cite{odlyzko}. The values of $I(j)$ for low $j$, shown in Fig.~\ref{sisjgue},
are in good agreement with those obtained from the ansatz (\ref{ij}).
Furthermore, we have computed for the same set of zeros the variance of the
total energy as a function of the number of occupied levels. Before
presenting the results we need to compute explicitly the third term in the
r.h.s. of Eq.(\ref{12}), denoted $V_{E,\zeta}^{(3)}$. An analogous
contribution for the number variance of the Riemann zeta function was
already considered in \cite{berry2}.

By comparing Eqs.(\ref{10p}) and (\ref{13}) one can see by analogy that
the period of the periodic orbits should be identified to the logarithm of
the prime numbers, $T_p = \log p$, and moreover $\hbar=1$. Then the rescaled
period is $\tau_p = \log p/\log(E_0/2\pi)$ and, from (\ref{13}), ${\hat A}_{p,r}
= - 2\pi\tau_p/(D p^{r/2})$. Then, as for any dynamical system, it
is expected that $K(\tau)$ behaves in a universal manner for $\tau>\tau^*$,
while ``small'' prime numbers should contribute with non-universal terms for
$\tau<\tau^*$. From the expression of $V_{E}^{(3)}$ in Eq.(\ref{12p}) and
using the above mentioned identifications we get
\begin{equation}\label{14}
V_{E,\zeta}^{(3)} = \frac{D^2 n^2}{2\pi^2} \sum_{r\tau_p<\tau^*} \frac{1}
    {r^2 p^r} \left\{\left[\cos(\pi n r\tau_p) - \frac{\sin
    (\pi n r\tau_p)}{\pi n r\tau_p}\right]^2 + \sin^2 (\pi n r\tau_p)\right\} \ .
\end{equation}
For the zeros we are considering we have $\tau_{\rm min}=\log
2/\log(E_0/2\pi) \approx 0.028$. We then choose $\tau^* = 0.2$ for the
numerical comparisons, which leads to a maximum prime number $p_{\rm max}=131$
in the sum in Eq.(\ref{14}).

\begin{figure}[thb]
\begin{center}
\mbox{\epsfig{file=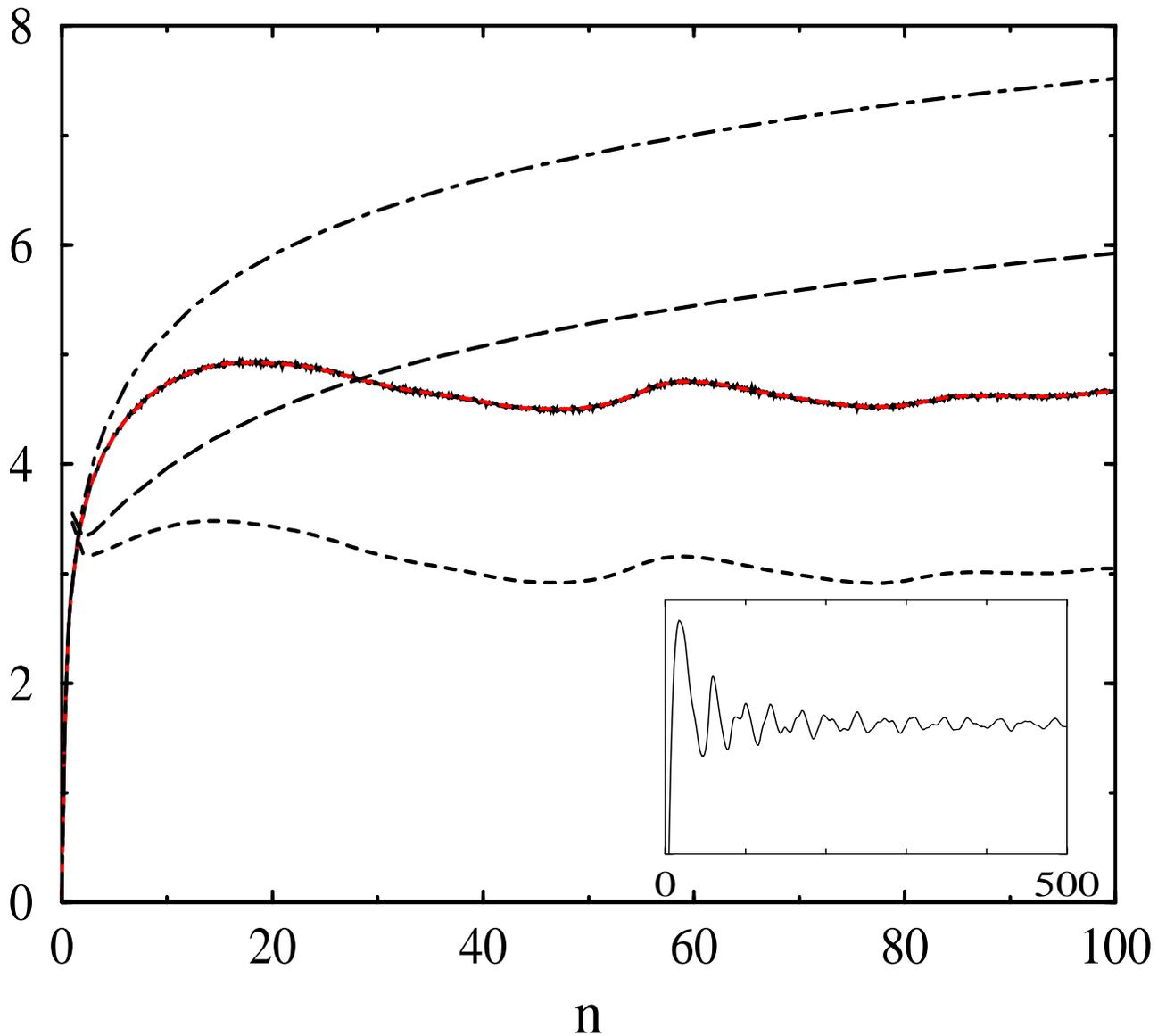,width=17cm,
bbllx=82pt, bblly=1pt, bburx=609pt,bbury=488pt,clip=}}
\end{center}
\caption{Normalized energy variances. Solid line: the normalized grand
canonical variance $V_E/(n^2/2\pi^2)$ for the zeros of the Riemann zeta
function. In this curve the numerical results as
well as the theoretical prediction are superimposed, and are almost
indistinguishable. The same curve is plotted in a wider range and on an
expanded scale in the inset. Dot-dashed line: random matrix result ($\beta=2$) for the
normalized grand canonical variance. Dashed line: the normalized canonical
variance $\Delta^{2}/(n^2/2\pi^2)$ obtained numerically from the zeros of
the Riemann zeta function. Long-dashed line: random matrix result
($\beta=2$, obtained using the ansatz) for the normalized canonical
variance. (See text for further explanation).}
\label{et2_fig4}
\end{figure}

In order to magnify the effect of non-universal corrections we have plotted
in Fig.~\ref{et2_fig4} the normalized quantity $V_{E}(n)/(n^2/2\pi^2)$ as a
function of $n$ (we set $D=1$). The solid line represents the theoretical
prediction (\ref{12}) with the third term of the r.h.s. given by
Eq.(\ref{14}). Superimposed to this curve the normalized variance computed
numerically from the zeros of the Riemann zeta function is also shown. The
agreement between the numerical results and the theoretical prediction is
remarkable and the two curves are almost indistinguishable. Also displayed
is the normalized GUE energy variance given by Eq.(\ref{8}). The saturation
effect with respect to random matrix theory is clearly visible for a number
of levels larger than $n\approx 5$. For higher values of $n$ some
oscillations are observed. These oscillations are well reproduced by the
prime-number contributions (\ref{14}). As predicted by the theory a damping
of the oscillations is observable and illustrated in the inset in a wider
range and on an expanded scale. The theory and numerical data are again in perfect
agreement. For large values of $n$ the curve tends to the constant $4.60$,
which is in very good agreement with the theoretical value given by
Eq.(\ref{12c}), $1-\log \tau_{\rm min} = 4.56$.

It is also interesting to display as a function of $n$ the behavior of the
canonical energy variance $\Delta^2$ for the zeros of the Riemann zeta
function. As was mentioned in the introduction, this variance is not a
two-point measure and therefore the semiclassical theories developed for the
form factor do not apply. We could again expect that, even in the
non-universal regime, for a large number of particles the canonical and
grand canonical variances are very closely related. Fig.~\ref{et2_fig4}
shows that indeed this is the case. The normalized energy variance $\Delta^2
/(n^2/2\pi^2)$ presents the same non-universal oscillations in the
saturation regime as $V_E$ (shifted however to a lower average value). The
non-universal features in $\Delta^2$ imply via Eq.(\ref{var}) their presence
in the autocorrelation function $I(j)$. This is confirmed by a study of
Odlyzko who has observed significant oscillatory deviations with respect to
random matrix theory for large values of $j$ \cite{odlyzko}, attributed to
effects due to primes.

As can be seen in Fig.~\ref{et2_fig4} the behavior of the two variances
$\Delta^2$ and $V_E$ is quite different for $n$ small. This difference is an
(uninteresting) manifestation of the discreteness of $n$ in the canonical
case.

\section{Concluding Remarks}

Assuming that the single particle spectrum is given by the eigenvalues of a
random matrix, we have determined the typical fluctuations of the total
energy of a system of non-interacting fermions. On the light of the original
applications of random matrix theory this approach may look at first sight
very unnatural. Indeed, these theories were originally applied to (nuclear)
many body systems in spectral regions and for properties unrelated to the
mean field, like for example neutron resonances of the compound nucleus.
Here we are somehow adopting the opposite point of view, namely to start
with an independent particle motion modeled by random matrix theory and
neglect completely residual interactions. Surprisingly, there are physical
situations for which this extreme view seems also to be fruitful, as should
become clear by the end of this section.

Two slightly different approaches have been investigated. In the first one,
denoted ``canonical'', we consider the fluctuations of a fixed number $n$ of
occupied levels. These fluctuations are shown to be directly related to the
autocorrelation function of consecutive spacings, for which we have proposed
an ansatz. The parameters entering the ansatz have been determined from the
asymptotic behavior of the fluctuations of the spacing variance $\sigma^2
(n)$. The canonical variance of the total energy is then computed, including
corrections up to order one. The leading-order term is found to be $n^2
\log n/(\beta \pi^2)$.

In the second -- ``grand canonical'' -- approach, the total energy variance
of the single particle levels contained in an interval of given length has
been considered. Contrary to the previous case, here the number of occupied
levels is not fixed. We have computed the grand canonical variance in the
limit of a large number of levels contained in the interval. Also an exact
result has been given for $\beta=2$. Both variances have the same leading-order
behavior, but the higher order corrections are different.

This difference in the higher order terms, and more generally the connection
between ``canonical'' and ``grand canonical'' quantities, have been studied
in detail and play a central role in the present study. Among the former we
have considered the spacing variance $\sigma^2$ and the energy variance
$\Delta^2$, while the number variance $\Sigma^2$ and the energy variance
$V_E$ belong to the latter. The connection (\ref{onesix}) between the
spacing and number variances has been essential in implementing the ansatz
(\ref{ij}). As already discussed in \cite{french}, Eq.(\ref{onesix}), though
asymptotic, remains a good approximation for any value of $n$. This is also
confirmed by our results. In fact, computing the difference to higher orders
we find
$$
\Sigma^2 (n) - \sigma^2 (n) = 1/6 +  {\cal O}(1/n^2) \ ,
$$
with the coefficient of the ${\cal O}(1/n^2)$ term being a small (of the
order $0.01$) constant for any $\beta$. In contrast, for the other two
quantities Eq.(\ref{6p}) is not a good approximation for small $n$.
Computing the difference to higher orders we find
\begin{equation} \label{c1}
\frac{V_E (n) - \Delta^2 (n)}{n^2} = \frac{1}{12} - \eta \frac{\log n}{n} -
\frac{1}{2} \left( C - \frac{1}{6} \right)\frac{1}{n} -
\frac{1}{3} \ (\eta - \lambda) \frac{\log n}{n^2} + {\cal O}(1/n^2) \ ,
\end{equation}
where the constants have been determined in Section II.A.

\begin{figure}[thb]
\begin{center}
\mbox{\epsfig{file=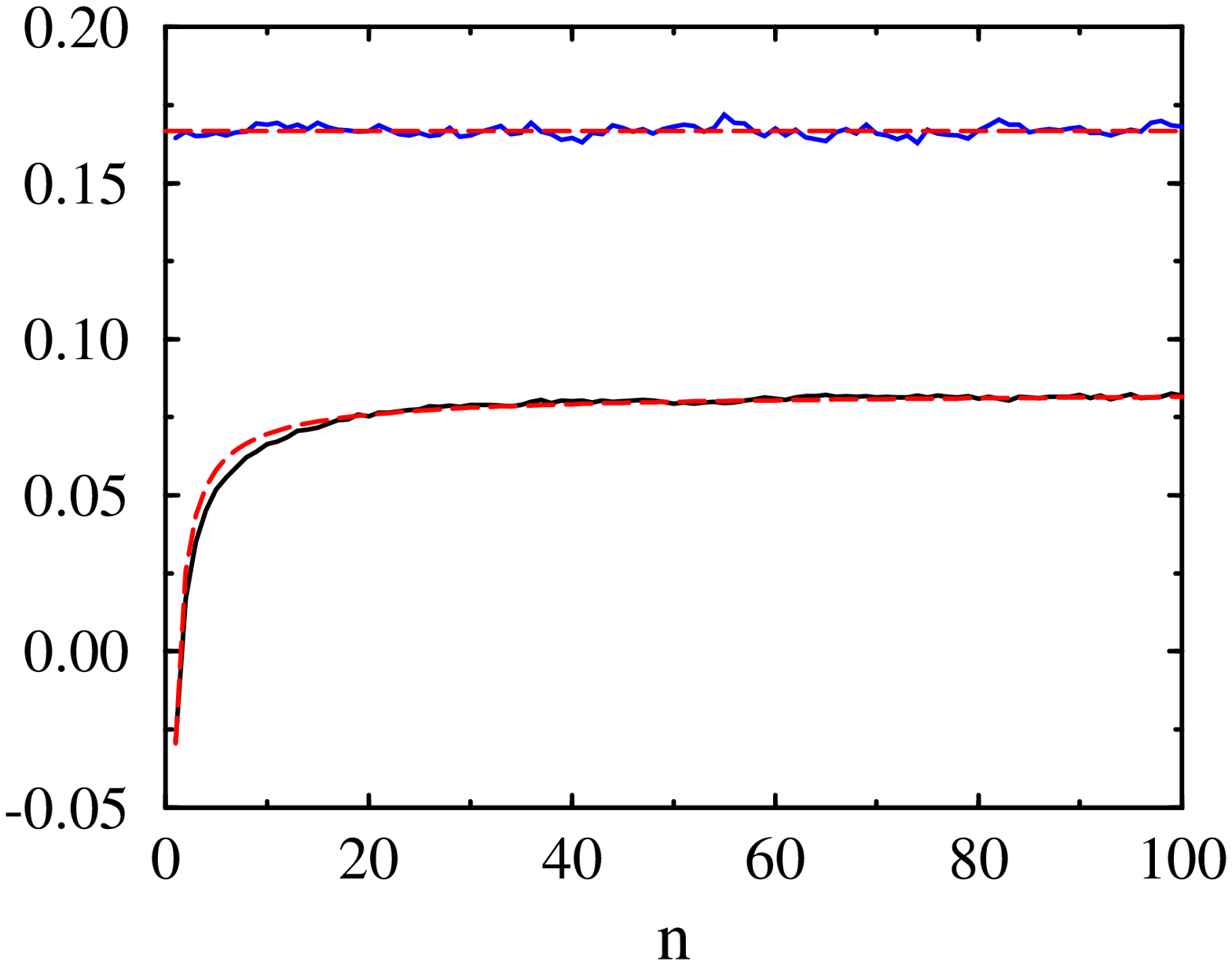,width=17cm,
bbllx=45pt, bblly=14pt, bburx=533pt,bbury=424pt,clip=}}
\end{center}
\caption{Comparison between canonical and grand canonical quantities. The
upper solid curve is the difference $\Sigma^2 - \sigma^2$ obtained
numerically from the zeros of the Riemann zeta function. The upper dashed 
horizontal line is at $1/6 \approx 0.166$. The lower solid curve is the
numerical result for the difference $(V_E- \Delta^2)/n^2$ for the same set
of zeros. The lower dashed curve displays Eq.(\ref{c1}), which tends to
$1/12 \approx 0.0833$ for large values of $n$. }
\label{et2_fig5}
\end{figure}

These relations between canonical and grand canonical quantities are those
predicted by random matrix theory, and thus applicable in principle only in
the universal regime (for example, Eq.(\ref{c1}) describes the difference
between the dot-dashed and the long dashed curves in Fig.~\ref{et2_fig4}).
But surprisingly we find that Eqs.(\ref{onesix}) and (\ref{c1}) {\sl always}
hold, {\sl irrespectively of the regime considered}. This ``universality in
the non-universal regime'' is illustrated in Fig.~\ref{et2_fig5} where both
differences for the zeros of the Riemann zeta function are plotted (as we
have mentioned before, a transition to a non-universal regime appears for
$n\ge 5$), and presumably it can be traced back to the validity of general
incompressibility conditions for a fluid \cite{dyson2,french}. An immediate
and interesting consequence of this observation is that the non-universal
corrections for canonical quantities (for which no explicit theory is
available) ought to be identical to those of grand canonical quantities. The
consequences of this remark and in particular the presence of non-universal
contributions in the behavior of the $I(j)$'s are under study.

The simplicity of periodic orbit theory in the interpretation of physical
phenomena has demonstrated its power in different branches of physics. A
well-known example is the prediction \cite{bb} and the experimental
confirmation in metallic clusters \cite{clusters} of supershell effects in
the density of states which can be basically understood in terms of a
beating produced by two short periodic orbits of a spherical cavity having
similar lengths. For irregularly shaped clusters, for which our model may be
a starting point, the possibility of having almost degenerate short periodic
orbits is not excluded. However it is unclear how robust this situation may
be under small perturbations, like for example the addition of particles to
the system. As already mentioned, similar and perhaps more promising
possibilities exist in the physics of quantum dots.

Other systems for which the present study may be relevant concern diffusive
systems in mesoscopic physics, like the motion of electrons in a disordered
piece of metal. Due to the presence of impurities, in this case the dynamics
of the electrons is not ballistic but rather described by a random walk. For
short times the non-universalities are not related to short periodic orbits
but to the return probability of a diffusive motion. This can be
incorporated in the short time behavior of the form factor \cite{ais}. In
particular it does not lead to the saturation effect discussed here and
presumably gives rise, for the variance of the total energy, to similar
effects as the ones observed for the number variance \cite{mo} which are
governed by the diffusion constant and the space dimensionality.

In nuclear physics there are some estimates of the fluctuations of the
binding energies due to non-systematic effects. When the macroscopic part as
well as shell corrections are substracted out from the measured total energy it
is found that the r.m.s. of the remaining fluctuations is about $0.5$ MeV
for heavy nuclei \cite{swiatecki}. Since the total binding energy for those
nuclei is about $1000$ MeV, the relative fluctuations due to non-systematic
effects are of order $R_{exp}=0.5 \times 10^{-3}$. With the present model,
since $\Delta \approx n \sqrt{\log n} /\pi$ for $\beta=1$ and $<X_t> \approx
n^2/2$, we obtain a theoretical estimate of the relative fluctuations
$R_{th} \approx 7\times 10^{-3}$ for $n \approx 200$ which overestimates the
experimental value by an order of magnitude. Before considering this
disagreement as significant obvious effects neglected here like the energy
variations of the mean level spacing should be taken into account.

\section*{Acknowledgments}
We are particularly indebted to A. Odlyzko for discussions and for providing
the set of zeros of the Riemann zeta function employed in our numerical
analysis. Discussions with W. Swiatecki provided some of the initial
motivations of the present work. M.J.S has been supported in part by a Thalmann
fellowship of the Universidad de Buenos Aires, Argentina.

\pagebreak

\renewcommand{\appendix}
       {\par
        \setcounter{section}{0}
        \setcounter{subsection}{0}
        \gdef\afterthesectionpunctdefault{:}
        \gdef\thesection{{APPENDIX}}
        \gdef\theequation{A\arabic{equation}}
        \gdef\lefteqno{\Alph{section}}
        \setcounter{equation}{0}}

\appendix
\section{A}
\renewcommand{\theequation}{\Alph{section}.\arabic{equation}} 

Given a stationary sequence of successive random single particles energies
$\ldots ,x_{i}, x_{i+1}, \dots $, take a point $O$ at random on the real
axis. It will lie between two levels, denoted by $x_{0}$ and $x_{1}$. Call
$s_{1}=x_{1}-x_{0}$ and $d$ the (random) distance between $x_1$ and $O$.
Fill the $n$ successive levels located just above $O$ and define the total
energy $X^{*}_t (n)$ with respect to $O$:
\begin{equation}
\label{et1}
X^{*}_{t}(n)= \sum_{i=2}^n (n-i+1) \; s_{i} \; + n \; {\it d} \equiv
\tilde{X}^{*}_t (n-1) + n \ d \; ,
\end{equation}
where $s_{i}=x_{i}-x_{i-1}$. By construction, $d$ is the product of two
independent random variables, $\epsilon$ and $s_1$, with $\epsilon$
distributed uniformly in the interval $[0,1]$ and $s_1$ with the nearest
neighbor spacing distribution of the original sequence. Using
$\langle \epsilon s_1 \rangle = \langle \epsilon \rangle \; \langle s_{1}
\rangle$, $\langle \epsilon \rangle = 1/2$ and  $\langle s_{1} \rangle= 1$
one has
\begin{equation}
\label{mean2}
\langle X^{*}_{t}(n)\rangle=\langle X_{t}(n-1)\rangle + \frac{n}{2} =
\frac{n^2}{2} \ .
\end{equation}

To compute the variance $\Delta^{* 2}(n)$ of $X^{*}_{t}(n)$, care must be
taken of the correlations between $s_1$ and the nearest neighbor spacings
above $s_1$. Using Eq.(\ref{sigma2}) one obtains
\begin{equation}
\label{term}
\langle s_{1}\; \tilde{X}_{t}(n-1) \rangle = \sum_{i=2}^n (n-i+1) \; 
\langle s_{1} \;s_{i} \rangle = \sum_{j=1}^{n-1} (n-j) \; I(j) = 
\left[ \sigma ^2 (n) - \; n \; 
\left( \langle s^2 \rangle - n \right) \right]/2 \ .
\end{equation}
Following the same steps as in section II, using the fact that $\epsilon$ is
uncorrelated with $s_1$ and $\tilde{X}_{t}$, using $\langle \epsilon^2 s_1^2
\rangle = \langle \epsilon^2 \rangle
\langle s_1^2 \rangle = \langle s^2 \rangle / 3$, and using Eq.(\ref{term})
one finally obtains
\begin{equation}
\label{varstar2}
\Delta^{* 2}(n)= \Delta^{2}(n-1) + \; \frac{n^2}{2} \; \left( \frac{1}{2} -
\frac{\langle s^2 \rangle}{3} \right) + \; \frac{n}{2}  \;  \sigma ^2 (n) \; .
\end{equation}

\section{B}
\renewcommand{\theequation}{\Alph{section}.\arabic{equation}} 

The energy variance is not independent of the reference point used to
measure the energy. In the grand canonical case, when an arbitrary point $a$
is used as reference point, $f(x)$ in Eq.(\ref{00}) should be taken as
\begin{equation} \label{last}
f(x) = \left\{ \begin{array}{ll}
     x - a \;\;\;\;\;\;\;  & \mbox{$-L\leq x \leq L$} \\
     0 \;\;\;\;\;\;\;  & \mbox{$|x|>L$} \ .
		\end{array} 
		\right. \nonumber  
\end{equation}
It follows from the definition of the energy variance (\ref{0}) and
(\ref{last}) that the variance $V_{E_0}$ with the reference point at the
center of the energy interval considered and the energy variance $V_{E_{a}}$
with the origin at $x=a$ are related through the equation
\begin{equation}\label{7p}
V_{E_{a}} (n) = V_{E_0} (n) + a^2 \, \Sigma^2 (L=n) \ ,
\end{equation}
where $\Sigma^2$ is the number variance. This is an exact relation valid for
any statistical sequence of energy levels (uncorrelated, GE, etc). 

When computing the energy variance we have used as reference point for the
energy the lower (or upper) end of the energy interval considered. Setting
$a=-L$ in Eq.(\ref{7p}), using Eq.(\ref{6}) and the asymptotic form of the
number variance (\ref{sig2}) we obtain for the variance with reference at
the center of the interval
\begin{equation}\label{7b}
 V_{E_0} (n) \approx \frac{\eta}{2} n^2 \log n + \frac{1}{2} \left(
 \frac{C}{2} - \eta \right) n^2 \ , 
\end{equation}
with the constants given in Section II.A. The leading order term in
Eq.(\ref{7b}) is twice smaller than the variance (\ref{6}) with the
reference point at the lower (or upper) end of the interval.

Because $ V_{E_0}$ and $\Sigma^2$ are positive definite, Eq.(\ref{7p}) shows,
incidentally, that the energy variance reaches a minimum when the reference
point is at the center of the interval.

\newpage

\newpage

\newpage

\newpage

\newpage

\begin{thebibliography}{99}
\bibitem{swiatecki} W. J. Swiatecki, {\sl Nucl. Phys. A} {\bf 574}, 233c (1994).
\bibitem{leshouches} ``Chaos and Quantum Physics'' edited by M.-J. Giannoni, A.
    Voros and J. Zinn-Justin, {\it Les Houches Session LII}, (North Holland,
    Amsterdam, 1991).
\bibitem{bohigas} O. Bohigas in Ref.\cite{leshouches}
\bibitem{politzer} H. Politzer, {\sl Phys. Rev. B} {\bf 40}, 11197 (1989).
\bibitem{diffusive} ``Mesoscopic Quantum Physics'' edited by E. Akkermans, G.
    Montambaux, J.-L. Pichard and J. Zinn-Justin, {\it Les Houches Session
    LXI}, (North Holland, Amsterdam, 1995).
\bibitem{mehta} M. L. Mehta, {\it Random Matrices}
(Academic Press, New York, 1991), 2nd ed. 
\bibitem{abram} M. Abramowitz and I. A. Stegun, {\it Handbook of Mathematical
Functions} (Dover, New York, 1970).
\bibitem{french} J. B. French, P. A. Mello and A. Pandey, {\sl Ann. Phys.} 
{\bf 113}, 277 (1978). T. A. Brody, J. Flores, J. B. French, P. A. Mello, A.
  Pandey and S. S. M. Wong, {\sl Rev. Mod. Phys.} {\bf53}, 385 (1981);
  J. B. French, V. K. B. Kota, A. Pandey and S. Tomsovic, {\sl Ann. Phys.}
{\bf 181}, 198 (1988); A. Pandey, unpublished.
\bibitem{odlyzko} A. M. Odlyzko, {\sl Math. Comp.} {\bf 48}, 273 (1987);
  AT \& T Report, 1989 (unpublished).
\bibitem{pandey} A. Pandey in ``Quantum Chaos and Statistical Nuclear
  Physics'' edited by T. H. Seligman and H. Nishioka, {\it Lectures Notes in
    Physics} {\bf 263}, (Springer Verlag, Berlin, 1986).
\bibitem{bhp} O. Bohigas, R. U. Haq and A. Pandey, {\sl Phys. Rev. Lett.}
  {\bf 54}, 1645 (1985).
\bibitem{dyson} F. J. Dyson and M. L. Mehta, {\sl J. Math. Phys.} {\bf 4}, 701
  (1963).
\bibitem{wilkinson} M. Wilkinson, private communication.
\bibitem{berry} M. V. Berry, {\sl Proc. R. Soc. London} {\bf A400}, 229
  (1985); see also M. V. Berry in Ref.\cite{leshouches}.
\bibitem{gutz} M. C. Gutzwiller,  {\it Chaos in Classical and Quantum
    Mechanics} (Springer Verlag, New York, 1990); see also M. C. Gutzwiller, in
    Ref.\cite{leshouches}.
\bibitem{ho} J. Hannay and A. M. Ozorio de Almeida, {\sl J. Phys. A} {\bf
    17}, 3429 (1984).
\bibitem{berry2} M. V. Berry, {\sl Nonlinearity} {\bf 1}, 399 (1988).
\bibitem{qdots} D. R. Stewart, D. Sprinzak, C. M. Marcus, C. I. Duru\"oz and
  J. S. Harris Jr, {\sl Science} {\bf 278}, 1784 (1997).
\bibitem{gm} H. L. Montgomery, {\sl Proc. Symp. Pure Math.} {\bf 24}, 181
  (1973); D. A. Goldston and H. L. Montgomery, {\sl Proc. Conf. at
    Oklahoma State Univ.}, edited by A. C. Adolphson {\sl et al}, 183 (1984).
\bibitem{bk} E. Bogomolny and J. Keating, {\sl Nonlinearity} {\bf 8}, 1115 
    (1995); {\sl ibid} {\bf 9}, 911 (1996); N. Katz and P. Sarnak, preprint
    Princeton 1997.
\bibitem{dyson2} F. J. Dyson, {\sl J. Math. Phys.} {\bf 3}, 166 (1962);
M. L. Mehta and F. J. Dyson, {\sl J. Math. Phys.} {\bf 4}, 713 (1963).
\bibitem{bb} R. Balian and C. Bloch, {\sl Ann. of Phys.} {\bf 69}, 76
  (1972); H. Nishioka, K. Hansen and B. R. Mottelson, {\sl Phys. Rev. B}
  {\bf 42}, 9377 (1990).
\bibitem{clusters} J. Pedersen, S. Bj{\o}rnholm, J. Borggreen, K. Hansen, T. P.
  Martin and H. D. Rasmussen, {\sl Nature} {\bf 353}, 733 (1991); T. P.
  Martin, S.  Bj{\o}rnholm, J. Borggreen, C. Br\'echignac, P. Cahuzac, K.
  Hansen, J. Pedersen, {\sl Chem. Phys. Lett.} {\bf 186}, 53 (1991).
\bibitem{ais} N. Argaman, Y. Imry and U. Smilansky, {\sl Phys. Rev. B}
  {\bf 47}, 4440 (1993).
\bibitem{mo}  G. Montambaux in ``Quantum Fluctuations'' edited by S. Reynaud,
  E. Giacobino and J. Zinn-Justin, {\it Les Houches Session LXIII}, (North
  Holland, Amsterdam, 1997).
\end{thebibliography}
\end{document}